\documentstyle[12pt]{article}

\begin{document}
\bibliographystyle{unsrt}

\begin{flushright} hep-ph/9410326

BA-94-56

UMD-PP-95-57

October 1994
\end{flushright}

\vspace{6mm}

\begin{center}

{\Large \bf Mass Matrix Textures From
 Superstring Inspired  SO(10) Models}\\ [6mm]

\vspace{6mm}
{\bf{K.S. Babu\footnote{Work supported by the Department of Energy
Grant \#DE-FG02-91ER406267}}}

{\it{Bartol Research Institute}}\\
{\it{University of Delaware}}

{\it{Newark, DE 19716}}

\vspace{3mm}

and

\vspace{3mm}

{\bf{R.N. Mohapatra\footnote{Work supported by the
 National Science Foundation Grant \#PHY-9119745}}}

{\it{ Department of Physics and Astronomy}}\\
{\it{University of Maryland}}\\

{\it{ College Park, MD 20742 }}

\end{center}

\vspace{4mm}

\begin{center}
{\bf Abstract}
\end{center}
\vspace{1mm}

We give a general prescription for deriving quark and lepton mass
matrices with ``texture'' zeros in the framework of superstring
inspired $SO(10)$ models. The key to our approach is a new way to
naturally implement the doublet--triplet splitting which enables
us to obtain symmetric quark and lepton mass matrices which have
different structures in the up and the down quark sectors.  We
illustrate our method by deriving the Georgi--Jarlskog texture
which has six predictions in the flavor sector, and then show how it
generalizes to other symmetric texture models.

\newpage

One of the major challenges in particle physics today is an
understanding of the spectrum of quarks and leptons.  In the standard
model, all the fermion masses and the mixing angles are arbitrary,
accounting for 13 of the 19 free parameters of the theory.  This
situation is clearly unsatisfactory and has been viewed by many as a
hint for physics beyond the standard model.  A more fundamental theory,
it is hoped, will reduce this arbitrariness, by providing a common
origin perhaps for several of these parameters.  In the absence of a
fundamental theory of fermion masses, it has been popular to assume certain
restricted forms for the quark and lepton mass matrices that results in
predictions for some of the observables.  The hope in
this approach is that if the
predictions of a particular mass matrix ansatz are borne out by
experiments, then a model that leads to a natural derivation of that
ansatz may provide the next step in our search for the fundamental
theory of nature.

The next step beyond the standard model may very well contain a
supersymmetric grand unified theory (SUSY GUT).  This speculation is
supported by the dramatically accurate unification of gauge
couplings that has been observed to occur [1] with supersymmetry around the
TeV scale following
the improved measurements of the low energy couplings at
LEP and SLC.  Superstring models,
which are candidate theories of unifying gravity with the strong and
electroweak forces, can lead to a variety of such SUSY GUT groups via an
appropriate compactification scheme.  It therefore appears to us to be a
very desirable program to see if a successful mass matrix ansatz can
receive a plausible derivation within a SUSY GUT model inspired by
superstrings.  This is what we take up in this paper.

There are a variety of mass matrix ansatze which can be
characterized by their ``texture'' zeros.  For increased
predictivity, it is usually assumed
that the matrices are symmetric.  A well--known example is the
Fritzsch ansatz [2] which has a
texture similar in form for the up and the down
quarks.  Such matrices are easily derived from GUTs such as $SO(10)$,
which has built--in left--right symmetry and up--down symmetry.  If the
recent CDF data [3] on the top quark, viz., $m_t = 174 \pm 17~GeV$, holds up,
it would appear that the Fritzsch ansatz would be ruled out even after
renormalization group corrections are taken into account [4].  Another
popular and predictive ansatz is the so--called Georgi--Jarlskog (GJ)
texture [5], which assumes symmetric quark and lepton mass matrices, which
however, have different forms in the up and down quark sectors.  Such
matrices, owing to their up--down asymmetry, are less trivial to derive
in the context of SUSY GUTs such as $SO(10)$.  In this letter,
we provide a prescription for deriving such {\it up--down asymmetric}
matrices in the context of superstring inspired $SO(10)$ models.  We
will illustrate our method by deriving the GJ texture, which has
been the subject of substantial improvement and polishing including
renormalization group effects in recent literature [6,7,8,9,10].
We will briefly
summarize the six predictions of the ansatz to show its consistency
with present data, especially with a heavy top as indicated by the
recent CDF results [3].  Then we show how
our method generalizes to other symmetric texture models.  In
particular, we show how it becomes possible to derive all of the five
symmetric models listed in Ref. [11].

SUSY $SO(10)$ seems to us to be the ideal setting for addressing the
quark and lepton masses.  All fermions of a family are unified to a
single irreducible representation of $SO(10)$,
which facilitates the generation of symmetric mass matrix
textures.  (This is not the case in $SU(5)$ GUTs, where the down quark
and the charged lepton mass matrices are not symmetric.
Higher symmetries such as $E_6$
invariably brings in exotic particles which can mix with the known
quarks and leptons, thus generally reducing the predictive power.)  The
emergence of the right--handed neutrino in $SO(10)$
leads to
small but non--zero neutrino masses via the
see--saw mechanism which may be highly desirable.
The vexing problem of doublet--triplet splitting
of SUSY GUTs also has an elegant
resolution in $SO(10)$.

It has been known for some time that conventional SUSY GUTs such as
$SO(10)$ with massless matter superfields belonging to the adjoint
representation (needed for gauge symmetry breaking)
can arise in the free fermionic formulation of
superstrings [12].  It is well--known that superstring theories
restrict considerably the number and
the nature of gauge multiplets that survive to low energies,
thereby reducing the
arbitrariness present in a general GUT.  They also provide naturally the
discrete symmetries which are often needed in restricting the texture of
quark and lepton mass matrices.  Specifically, we are encouraged by the
recent works of Choudhuri, Chung and Lykken [13] and Cleaver [14] who have
constructed explicit
$SO(10)$ models with adjoint scalars at the Kac-Moody level
of two.  These authors also classify the allowed representations that
emerge as massless chiral multiplets below the Planck scale at
the level two construction.  While
there can be any number of vectors ({\bf 10}), spinors
$(\overline{\bf 16} + {\bf 16})$ and gauge singlets, the number of
adjoints is restricted to be at most 2.
Similarly, no more than one {\bf 54} can remain
light, although no explicit example with any {\bf 54}
has been constructed so far.

We shall be guided in our derivation of
the mass matrix ``textures'' by the superstring constraints
listed above.  Specifically, the Higgs representations
that we shall use for symmetry breaking and for fermion mass
generation will be a spinorial $\overline{\bf 16}+{\bf 16}$, a
pair of {\bf 10}, two adjoint {\bf 45} and a few singlets.
In order to make a realistic and predictive model for quarks and leptons
at low energies, we shall impose the following requirements:
(i) there must be a consistent mechanism to break the SO(10) symmetry
down to the standard model at the GUT scale;
(ii) the light particle spectrum
of the theory must be such that it preserves the successful
prediction for sin$^2\theta_W$; (iii) the doublet--triplet splitting
should be achieved naturally (i.e., without fine--tuning)
in such a way
that only one pair of Higgs doublet remains light (to be identified
with $H_u$ and $H_d$ of MSSM); and finally,
(iv) the same symmetry that helps satisfy
the above requirements must provide an interesting texture for fermion mass
matrices.

Let us first briefly summarize the predictions of the Georgi--Jarlskog
ansatz.  It assumes the following form
for the up quark, down quark and the charged lepton mass matrices at
the GUT scale:
\begin{eqnarray}
M_u~=\left(\begin{array}{ccc}
0 & a & 0\\
a & 0 & b\\
0 & b & c \end{array}\right);~
M_d~=\left(\begin{array}{ccc}
0 & d e^{i\phi} & 0\\
d e^{-i\phi} & f & 0\\
0 & 0 & g\end{array} \right);~
M_\ell~=\left(\begin{array}{ccc}
0 & d & 0\\
d & -3f & 0\\
0 & 0 & g \end{array}\right).
\end{eqnarray}
\noindent There are 7 parameters in all to fit the 13
observables (9 masses, 3 mixing angles and one CP phase), thereby
resulting in six predictions.  Three of these predictions are
the $b,s$ and $d$--quark masses:
\begin{eqnarray}
m_b = \eta^{-1}_{b/\tau} m_\tau; ~ {{m_d/m_s}\over {(1-m_d/m_s)^2}} =
9 {{m_e/m_\mu} \over {(1-m_e/m_\mu)^2}};~(m_s-m_d) = {1 \over 3}
\eta_{s/\mu}^{-1}(m_\mu-m_e)~.
\end{eqnarray}
The other three predictions are for the quark mixing angles and the CP
phase $J$:
$$
|V_{cb}| = \eta_{KM}^{-1}\eta_{u/t}^{1/2}\sqrt{{{m_c}\over {m_t}}}~;
{}~~~~~~~~{{|V_{ub}|} \over {|V_{cb}|}} = \sqrt{{{m_u}\over {m_c}}} ~;
$$
\begin{equation}
J = \eta_{KM}^{-2} \eta_{u/t} \sqrt{{{m_d}\over {m_s}}}
\sqrt{{{m_c}\over {m_t}}}\sqrt{{{m_u}\over {m_t}}}
\left[1-{1 \over 4} \left(\sqrt{{{m_u}\over {m_c}}}
\sqrt{{{m_s}\over {m_d}}}+\sqrt{{{m_c}\over {m_u}}}
\sqrt{{{m_d}\over {m_s}}}-
\sqrt{{{m_c}\over {m_u}}}\sqrt{{{m_s}\over {m_d}}}|V_{us}|^2\right)^2
\right]^{\frac {1} {2}}
\nonumber \\
\end{equation}
Here the $\eta$'s are renormalization factors for the various
parameters in going from the low energy scale to the GUT scale.
If the bottom--quark Yukawa coupling $h_b$ is much smaller than the
top Yukawa coupling $h_t$, (corresponding to tan$\beta
\stackrel{_<}{_\sim} 10$ or so)
these RGE factors can be expressed analytically as
\begin{eqnarray}
\eta_{KM} &=& \eta_{d/b}=\left(1-{{Y_t}\over {Y_f}}\right)^{{1 \over 12}};~
\eta_{u/t} = \left(1-{{Y_t}\over {Y_f}}\right)^{{1 \over 4}};
\eta_{s/\mu} = \left({{\alpha_1}\over {\alpha_G}}\right)^{-10/99}
\left({{\alpha_3}\over {\alpha_G}}\right)^{-8/9}; \nonumber \\
\eta_{b/\tau} &=& \left({{\alpha_1}\over {\alpha_G}}\right)^{-10/99}
\left({{\alpha_3}\over {\alpha_G}}\right)^{-8/9}\left(1-{{Y_t}\over
{Y_f}}\right)^{-1/12}~.
\end{eqnarray}
Here
$\alpha_G$ is the unified gauge coupling strength,
$Y_t = h_t^2$ at the weak scale
and $Y_f$ is the fixed
point value of $Y_t$.  $Y_t$ cannot strictly be equal to $Y_f$, since
that would correspond to infinite $Y_t$ at the GUT scale.
If we demand that $Y_t \stackrel{_<}{_\sim} 4$
at GUT scale, $Y_t/Y_f$ can at
most be 0.98.  The renormalization factors in Eq. (4)
are all well--behaved
even when $Y_t$ differs from $Y_f$ by only 2\% due to the small
exponents.

In the three Yukawa unification scheme
($h_t=h_b=h_\tau$ at the GUT scale), the exponents $(1/12, 1/12, 1/4)$
corresponding to $(\eta_{KM}, \eta_{d/b}, \eta_{u/t})$ in Eq. (4)
will be replaced by $(1/7, 2/7, 2/7)$; $\eta_{s/\mu}$ will remain
unchanged, while $\eta_{b/\tau}$ is modified to
\begin{equation}
\eta_{b/\tau} = \left({{\alpha_1}\over{\alpha_G}}\right)^{-19/376}
\left({{\alpha_3}\over {\alpha_G}}\right)^{-2/9} \left({{Y_t}\over
{Y_\tau}}\right)^{-3/8}\left(1-{{Y_t}\over {Y_f}}\right)^{-1/14}~.
\end{equation}
The $\eta$ factors given in Eqs. (2)-(5) correspond to running
the parameters from the weak scale (taken here to be $m_t$
which is also assumed to be the SUSY threshold)
up to $M_{GUT}$.  For the
light fermion masses there is QCD and QED running factors below $m_t$ as
well.  These are obtained numerically using
three loop QCD and
one--loop QED $\beta$ and $\gamma$
functions.  Corresponding to $\alpha_s({M_Z}) = 0.12$ and
$\alpha^{-1}(M_Z) =127.9$,  these factors are
$(\eta_u,\eta_{d,s},\eta_c,\eta_b,\eta_{e,\mu},\eta_\tau) =
(0.401,0.404,0.460,0.646,0.982,0.984)$.
If $\alpha_s(M_Z)=0.125$ is chosen these
factors become $(0.356,0.358,0.422,0.630,0.982,0.984)$.
Using $\alpha_s(M_Z) = 0.12, m_c(m_c) = 1.27~ GeV, m_u(1~GeV) = 5.1~ MeV,$
$|V_{us}| = 0.22$, $m_t^{\rm phys} = 174~GeV$ and
$Y_t/Y_f = 0.98$ as input values, we obtain
$m_d(1~GeV) = 7.7~MeV$, $m_s(1~GeV) = 193~MeV, m_b(m_b)=4.26~GeV,
|V_{cb}| = 0.050$,
$|V_{ub}|/|V_{cb}|
= 0.059$, $J=2.96 \times 10^{-5}$.
The value of $|V_{cb}|$ corresponding to $\alpha_s(M_Z) =
0.125, m_c(m_c) = 1.22~GeV, m_t^{\rm phys} = 190~GeV$ is
$|V_{cb}| = 0.045$.  In the case of three Yukawa unification,
$h_t=h_b=h_\tau$, $|V_{cb}|$ is
unrenormalized and results in a larger value, which is therefore
disfavored.  Note that all the predictions of the GJ ansatz
are presently in good agreement with experiments, especially for the
case of small tan$\beta$.

We now turn to the derivation of Eq. (1), which is our main result.
Notice that the up and the down quark mass matrices in Eq. (1) have a
very asymmetric structure.  We will show how to generate such an
up--down asymmetry within $SO(10)$.
First we observe that all elements of
the fermion mass matrices  must arise from effective scalar operators that
transform as {\bf 10}, {\bf 120} or $\overline{\bf 126}$ of
$SO(10)$, each of which can couple to the fermion bilinears.
The {\bf 10} and $\overline{\bf 126}$ operators result is symmetric mass
matrices.  In order to obtain an up--down asymmetry,
the effective {\bf 10} or $\overline{\bf 126}$ operator that generates the
(23) element of $M_u$ must develop an electroweak VEV only along the
up (and not along the down) direction.  Similarly, the $\overline{\bf 126}$
operator that induces the (22) element of $M_d$ and $M_l$ must develop an
electroweak VEV only along the down direction.  Now, if the successful
prediction of sin$^2\theta_W$ of SUSY GUTs is to be preserved,
only one pair of Higgs(ino) doublets can remain light
($H_u$ and $H_d$ of MSSM).  These
are the only doublets that acquire electroweak VEVs.  There could be
several doublet
Higgs(ino)s at the GUT scale which mix with one another, but
in such a way that only one pair of them remains light [15].
The texture of the
mass matrices in Eq. (1) then requires that the Higgs doublet coupling to
the (23) entry in $M_u$ should have an admixture of $H_u$, but not of
$H_d$ and similarly, the Higgs doublet generating (22) entry of
$M_d$ and $M_l$ should have an admixture of $H_d$, but not $H_u$.  This
leads to the important observation that the texture
zeros of the fermion mass
matrices can be ensured only if the Higgs(ino) mass matrix has its own
texture zeros.  Furthermore, it requires that
the Higgs(ino) mass matrix be
asymmetric, otherwise an up--down asymmetry cannot be induced.

Which linear combination of up--type and down--type Higgs(ino) doublets
remains light is of course related to the question of doublet--triplet
splitting of SUSY GUTs.  The simplest way to achieves a natural
doublet--triplet splitting in $SO(10)$ is via the
Dimopoulos--Wilczek
mechanism[16].  This is achieved by the coupling
$H_1AH_2$ where $H_1$ and $H_2$ belong to the
{\bf 10} of $SO(10)$, while $A$ is the adjoint {\bf 45}.  This term
gives masses to the color triplets in {\bf 10}'s but not to the
$SU(2)$ doublets if the VEV of the adjoint $A$ is chosen to be along
the $(B-L)$ direction: $\langle A \rangle =
diag.(a,a,a,0,0)\times i\tau_2$. In realistic SO(10) models, this pattern
of vev's is not stable and new techniques are needed to make this
method useful [17,18].
Furthermore, it leads to up-down symmetric
mass matrix textures for the fermions,
which are not useful for our purpose.

In order to generate an asymmetry in the Higgs(ino) mass matrix
we propose to mix the {\bf 10}-plets ($H_1$ and $H_2$)
with the spinorial {\bf 16} +
$\overline{\bf 16}$ ($\psi_H+\overline{\psi}_H$),
which are needed anyway for symmetry breaking.
Each {\bf 10} contains in it
one up--type and one down--type
Higgs doublet, $\psi_H$ contains a down--type doublet, while
$\overline{\psi}_H$ contains an up--type doublet.  We have therefore a
total of three up--type and three down--type Higgs doublets.
The {\bf 10}'s also
contain a color triplet--antitriplet pair, $\psi_H$ has a color
antitriplet while $\overline{\psi}_H$ has the color triplet.
The superpotential involving the $SU(2)$ doublets and the color triplets
is assumed to be
\begin{equation}
W_{DT} = \lambda_1 \psi_H\psi_HH_1 + \lambda_2 \overline{\psi}_H
\overline{\psi}_H H_2 + H_1AH_2~.
\end{equation}
This is the most general superpotential relevant for the
doublet--triplet splitting compatible with the set of
discrete symmetries given in Table. 1.
Eq. (6) leads to the following mass matrices for the Higgs(ino) doublets
and color triplets
(written in the basis where the rows correspond to $(\psi_{H_d},
H_{2d}, H_{1d})$ and the columns stand for $(\overline{\psi}_{H_u},
H_{1u}, H_{2u})$ for the doublets and similarly for the color triplets)
\begin{eqnarray}
M_D = \left(\matrix{0 & \lambda_1 v_R & 0 \cr
\lambda_2 v_R & 0 & 0 \cr
0 & 0 & 0}\right);~~M_T = \left(\matrix{0 & \lambda_1 v_R & 0 \cr
\lambda_2 v_R & \lambda_3 a & 0 \cr 0 & 0 & -\lambda_3 a}\right)~.
\end{eqnarray}
Here $\langle \psi_H \rangle = \langle \overline{\psi}_H \rangle
= v_R$ and $\langle A \rangle = a$.  This gives GUT scale Dirac masses
to all three
triplet--antitriplet pairs.  Two pairs of Higgs(ino)
doublets become superheavy, leaving one pair of
doublets light.  The $H_u$ and $H_d$ fields
of MSSM are easily identified as
$H_u = H_{2u}$ and $H_d = H_{1d}$.
Note that the desired asymmetry has been achieved,
$H_1$ will acquire a VEV only along the down
direction ($\langle H_{1d} \rangle  \neq 0, \langle H_{1u}\rangle = 0$),
while $H_2$ will acquire a VEV only along the up direction
($\langle H_{2d} \rangle = 0, \langle H_{2u} \rangle \neq 0$).  This
serves as the first step in our derivation of the GJ ansatz.  In this
scheme, tan$\beta \neq m_t/m_b$ owing to the ({\bf 10}-{\bf 16}) mixing.

Let us now turn to the symmetry breaking sector.
Nonrenormalizable operators suppressed by appropriate powers of inverse
Planck mass, which are expected to arise in superstring theories, will
play a crucial role in our analysis.  In addition, we shall make use of
a few gauge singlet superfields
to generate the GUT scale starting
from the Planck scale and the SUSY breaking scale $m_0 \sim 1~TeV$.  To
see how this works,
consider a nonrenormalizable superpotential involving a
singlet superfield
\begin{equation}
W = \lambda {{\phi^n} \over {M_{Pl}^{n-3}}}~
\end{equation}
where $n$ is an integer.
The corresponding scalar potential, including soft supersymmetry breaking
terms is given by
\begin{equation}
V=m_0^2 |\phi|^2 + \left|{{n\lambda \phi^{n-1}}\over {M_{Pl}}^{n-3}}
\right|^2 + m_0A_n \left[{{\lambda \phi^n}\over {M_{Pl}}^{n-3}}+H.C.
\right]~
\end{equation}
In the supersymmetric limit, the potential has a unique minimum given by
$\langle \phi \rangle = 0$.  But including SUSY breaking, the potential
develops two other minima given by
\begin{equation}
\langle |\phi| \rangle = M_{Pl}\left[{1 \over 2} \left|{{A_n m_0}\over
{\lambda M_{Pl}
n(n-1)}}
\right| \left(1\pm \sqrt{1-{{4(n-1)}\over {A_n^2}}}\right)
\right]^{1 \over {n-2}}~.
\end{equation}
For $n=8,9,10$, this leads to
$\langle |\phi| \rangle \sim 10^{16}~GeV$ which is near the GUT scale.  The
existence of these symmetry breaking minima
requires that $A_n^2 \ge 4(n-1)$ (see Eq. (10)).  We have
investigated if this constraint is satisfied automatically in $N=1$
supergravity models with hidden sector SUSY breaking $a~ la$ Polony.
We found $|A_n| = |n-\sqrt{3}|$ in this scheme, which satisfies
the constraint for $n \ge 7$.

Let us now display the part of the superpotential that breaks the
gauge symmetry down to the standard model.  A minimum of
one adjoint {\bf 45} and a {\bf 16}+$\overline{\bf 16}$ pair Higgs
superfields are required for this [17,18].  In order to
generate the GJ texture, we shall also need a second adjoint
$A^\prime$. $A$, which has a VEV along the $(B-L)$ direction,
$\langle A \rangle= diag.(a,a,a,0,0)\times i\tau_2$, is
responsible for the doublet--triplet splitting.  $A^\prime$ develops a
VEV along the $I_{3R}$ direction, $\langle A^\prime \rangle =
diag.(0,0,0,a^\prime,a^\prime)\times i\tau_2$.
In addition, we shall need a few
singlets, in order to generate the GUT scale as discussed above, as well
as for the fermion mass matrix texture.  These singlets are denoted by
$X,Y,Z,\overline{Z}$ and $S_{1,2,3}$, all of which develop GUT scale
VEVs except for the field $X$ which has zero VEV.  The superpotential
which achieves the symmetry breaking and the desired  VEV structure
is given by
\begin{eqnarray}
W_{sym}= X(\bar{\psi}_H\psi_H~-~Y^2)~+ A^2 Z^2+
A^2\bar{Z}^2 + A^4 + A^2 A^{{\prime}2}+
A^{\prime^2}Z^2 + A^{\prime^2}\overline{Z}^2+ \nonumber \\
A^{\prime^4}
+ (Z\overline{Z})^2 + Z^4 + \overline{Z}^4+
M \overline{Z}Z +S_1^8+S_2^8+S_1^2S_2^4S_3^4+X^8A^2 + ...
\end{eqnarray}
This is the most general superpotential up to relevant orders consistent
with the discrete symmetries given in Table 1.  We should mention here
that we have not tried to be economical with the choice of the discrete
symmetry.  Rather, we have identified the maximal symmetry of the
desired superpotential in Table 1.  The full symmetry of Table 1 may not
be essential to achieve GJ texture.
It is understood that
in Eq. (11), each term is suppressed by the dimensionally appropriate
powers of $M_{Pl}$.  Terms such as $A^4, A^2A^{\prime^2}$ in Eq. (11)
stand for more than one possible contraction of indices.  If the mass
parameter $M$ is chosen to be $M_{GUT}^2/M_{Pl}$, $\langle A \rangle,
\langle A^\prime \rangle \sim M_{GUT}$ will be generated.  The
superpotential gives rise to the VEVs of $A$ and $A^{\prime}$ in the
desired $(B-L)$ and $I_{3R}$ directions respectively.
Except for the question
of linking the $(A,A^\prime)$ sector to the $(\psi_H,\overline{\psi}_H)$
sector (this issue will be addressed shortly),
which is necessary to avoid pseudogoldstone bosons, the symmetry
breaking down to the standard model has now been achieved.  As for the
corrections to the doublet Higgs(ino)
mass matrix, the leading order terms are
$H_1H_2Y^4/M_{Pl}^3$ and $H_1H_2(\psi_H\overline{\psi}_H)^2/M_{Pl}^3$.
These will induce an effective $\mu$ parameter for $H_u$ and $H_d$ of MSSM.
Demanding that the $\mu$ parameter is $\stackrel{_<}{_\sim} 1~TeV$,
we find that
$\langle \psi_H \rangle, \langle Y \rangle \sim 10^{15}~GeV$, which is
satisfactorily close to the GUT scale.  Terms such as $\psi_H\psi_HH_2$
that can also contribute to the $\mu$ parameter do not arise until very
high order.

We now proceed to the derivation of the GJ texture.
The relevant superpotential involving the matter fields
consistent with the symmetries of Table 1 is
\begin{eqnarray}
W_{Yuk} & = & M^{-1}\left(h_{33}\psi_3\psi_3 H_1 S_1 +h^{\prime}_{33}
\psi_3\psi_3 H_2 S_2 + h_{23}\psi_2\psi_3H_2 S_3\right) + \nonumber \\
& ~& M^{-2}h_{22}
\psi_2\psi_2 H_1 A A^{\prime}
+M^{-3}\left(h_{12}\psi_1\psi_2H_1S^3_2
+h^{\prime}_{12}\psi_1\psi_2H_2S^3_1\right)~.
\end{eqnarray}
Here $M$ is not necessarily the Planck mass, Eq. (12) could arise by
integrating out vector--like fermions such as $\overline{\bf 16}$ +
{\bf 16} which have an intermediate mass between the GUT and the
Planck scale.  All parameters in Eq. (12) can be made real by field
redefinitions, but the VEVs of $A,A^\prime$ are complex in general, so
one unremovable phase will reappear after symmetry breaking
as in Eq. (1).

The $h_{22}$ term in Eq. (13) has quite interesting properties, which
needs a bit of explanation.  In
general, the $H_1AA^\prime$ can give rise to effective {\bf 10}
as well as $\overline{\bf 126}$ operators.
The GJ texture requires that only the
$\overline{\bf 126}$ contributes.  The interesting feature of this term
with $\langle A \rangle$ along $(B-L)$ direction and $\langle A^\prime
\rangle$ along $I_{3R}$ direction is that only the $\overline{\bf 126}$
operator contributes to the mass matrices.  Effective {\bf 10} can arise
in three different ways, all of which vanish at the minimum, due to the
orthogonality of $\langle A \rangle$ and $\langle A^\prime \rangle$.
To see this, we first observe that Tr$(AA^\prime) = 0$, which sets one
such contribution to zero.  Other terms of the type $A_{ab}
A^\prime_{bc}H_{1c}$  and $A^\prime_{ab}A_{bc}H_{1c}$
also vanish at the minimum.  One is left with
the term $A_{ab}A^\prime_{cd}H_e$, which is precisely the
$\overline{\bf 126}$ contribution.  This does not vanish and as a result,
we are able to reproduce the relative factor of $-3$ in the (22)
entry of $M_d$ and $M_l$.

If all the Yukawa coupling parameters in Eq. (12) are chosen
to be of order one, the hierarchy in the masses can be explained purely
by the ratio $\langle S_i \rangle/M \equiv \epsilon$.  Except for $S_2$,
which generates the top quark mass, we will choose these ratios to be in
the range $1/10$ to $1/30$.
As noted already, the doublet--triplet splitting mechanism
implies that $H_{1d} = H_d$ and $H_{2u} = H_u$.  Making this
identification, we see that
below the GUT scale, Eq. (12) results in the following effective
Yukawa superpotential (with redefined $h_{ij}$):
\begin{eqnarray}
W^{eff}_{Yuk}&=&h^{\prime}_{33}Q_3H_u u^c_3 +\epsilon h_{33}(Q_3H_d d^c_3
+L_3H_d e^c_3)+h_{23}\epsilon(Q_2H_u u^c_3+Q_3H_u u^c_2) + \nonumber \\
&~&h_{22}\epsilon^2
(Q_2H_d d^c_2 -3 L_2 H_d e^c_2) + h^{\prime}_{12}
\epsilon^3 (Q_1H_u u^c_2+Q_2H_u u^c_1) + \nonumber \\
&~& h_{12}\epsilon^3 (Q_1 H_d d^c_2
+ Q_2 H_d d^c_1 +L_1 H_d e^c_2 + L_2 H_d e^c_1)~.
\end{eqnarray}
The desired GJ texture follows from this after electro-weak symmetry
breaking.  It is easy to see that the texture zeros of the mass
matrices are protected to very high order by the discrete symmetry.

Let us now turn briefly to the neutrino sector.  The Dirac neutrino
matrix is identical to the up quark matrix in the model.  As for the
right--handed neutrino Majorana mass matrix,
we have no more freedom to
choose its form since its discrete symmetry assignment has already been
fixed in the process of deriving the GJ ansatz.
The allowed right--handed neutrino mass terms are
\begin{equation}
W_{\nu_R} = \left(M^{-3} \psi_3 \psi_3 S_1A+M^{-2} \psi_2\psi_2 A^\prime
+ M^{-10} \psi_1\psi_1A^{\prime^3}S_2^6\right)\overline{\psi}_H
\overline{\psi}_H~.
\end{equation}
The resulting $\nu_R$ mass matrix is nonsingular, and thus generates
small neutrino masses for $\nu_e,\nu_\mu$ and $\nu_\tau$.  Note that
although the last term in Eq. (14) has higher inverse powers of $M$, it
is compensated largely by the VEV of $S_2$, which cannot be too much
below $M$ as it generates the top quark mass.  Thus the neutrino
spectrum is similar to what is expected in non--SUSY $SO(10)$ models
with an intermediate scale.  Such a spectrum is known to be capable of
resolving the solar neutrino puzzle via the MSW mechanism while leaving
the $\nu_\tau$ mass in the cosmologically significant multi--eV range.
It is interesting that the superstring
constraint on the Higgs spectrum is what is responsible for the
weakening of
the see--saw suppressions somewhat.

Let us turn now to the question of linking the $(A,A^\prime)$ and
the $(\overline{\psi}_H, \psi_H)$ sectors.
If the two sectors were not linked, there will
be pseudogoldstones belonging to
a {\bf 10} + $\overline{\bf 10}$ of $SU(5)$.
Although the successful prediction of
sin$^2\theta_W$ will be unaffected,  since
they form complete multiplets of $SU(5)$, they do upset the
prediction for $m_b$.  In order to make them superheavy, one
cannot link the two sectors directly, as that would upset the VEV of
$A$.  The simplest way to achieve this is to assume a term
Tr$(AA^\prime A^{\prime \prime})$ along with $A^{\prime \prime}
\overline{\psi}_H\psi_H$ in the superpotential.  Here $A^{\prime \prime}$
is another adjoint.  The Tr$(AA^\prime A^{\prime \prime})$ term, due to
its complete anti-symmetry,
does not upset the VEV structure of $A,A^\prime$, yet
it makes all the pseudogoldstones superheavy.
The introduction of a third
adjoint would appear to be in conflict with the superstring
constraints, however, $A^{\prime \prime}$ need not
survive below the Planck scale, in which case there is no contradiction.
To be specific, let us add the following terms to the superpotential:
\begin{equation}
W^\prime = \overline{\psi}_H \psi_H A^{\prime \prime}P + AA^\prime
A^{\prime \prime} Q + (A^{\prime \prime}Q)^2 + Q\overline{Q} +
(Q\overline{Q})^2 + (P\overline{Q})^8
\end{equation}
where $P,Q,\overline{Q}$ are gauge singlets.
These terms are clearly consistent with
the discrete symmetries of Table 1.  $W^\prime$ also has a $Z_n$ symmetry
under which $(A^{\prime \prime}, \overline{Q})$ and $(P,Q)$ have opposite
charges.  This superpotential admits $\langle Q \rangle \sim M_{Pl}$, so
that $A^{\prime \prime}$ has a mass of order $M_{Pl}$.  $P$ gets a VEV
of order $M_{GUT}$ from the last term in Eq. (15).
The coupling of $A^{\prime \prime}$  to
the other fields make all
pseudogoldstones heavy, of order $M_{GUT}^2/M_{Pl}$.  $A^{\prime \prime}$
will receive an induced VEV, but it is of order $M_{GUT}^3/M^2_{Pl}$.  The
only effect of $A^{\prime \prime}$ on the fermion mass matrices is to
give a tiny correction to the (22) element of $M_{d,l}$,
via the term $\psi_2\psi_2 H_1 A^{\prime \prime}Q$, which is about
$10^{-3}$ times smaller than the leading term.

Let us finally show how the method developed here facilitates the
derivation of other symmetric mass matrix textures.  Take for example,
Model (4) of Ref. (11), which is obtained by adding a (22) entry in
$M_u$ of Eq. (1).  Such a texture
follows readily in our scheme by a new Yukawa
term $\psi_2 \psi_2 H_2 S$ with $S$ a gauge singlet carrying discrete
charge of $(4,4,0,2)$.  All the five symmetric texture models
of Ref. (11) can be derived in an analogous fashion.  Our method can be
applied to derive asymmetric texture models as well in the context of
$SO(10)$ [19].

Let us conclude by observing some interesting variations of the
doublet--triplet splitting scheme (Eqs. (6)-(7)).
One could add to Eq. (6) direct mass terms
$(M_\psi\psi_H\overline{\psi}_H + M_1 H_1^2 + M_2H_2^2)$ with
$M_\psi M_1M_2=0$ to ensure a light doublet.  If $M_1=0$, the MSSM
fields $H_u$ and $H_d$ are $H_u \propto (M_\psi M_2 H_{1u} +
\lambda_1 \lambda_2 v_R^2H_{2u} - \lambda_1 v_R M_2 \overline{\psi}_{H_u})$
and $H_d=H_{1d}$.  Thus $\langle H_{1u} \rangle, \langle
H_{1d} \rangle, \langle H_{2u} \rangle, \langle \overline{\psi}_{H_u}
\rangle \neq 0$, while $\langle H_{2d} \rangle =\langle
\overline{\psi}_{H_d} \rangle = 0$.  Such a spectrum enables one to
use Yukawa couplings to the $H_1$ field to induce some common elements
in $M_{u,d,l}$ while generating an up--down asymmetry via the couplings
of $H_2$ and $\overline{\psi}_{H_u}$.  If $M_\psi = 0, M_1M_2 \neq 0$,
then $H_d = {\rm cos}\theta H_{1d}+{\rm sin}\theta \psi_{H_d}$ and
$H_u = {\rm cos}\theta^\prime H_{2u}+{\rm sin}\theta^\prime
\overline{\psi}_{H_u}$, with $\theta,\theta^\prime$ Higgs(ino) mixing
angles.  The up--down asymmetry is similar to Eq. (7), but now there
is the freedom of involving the $\psi_{H_d},\overline{\psi}_{H_u}$
fields in the fermion mass generation.  Details of these and application
to other textures will be the subject of a forthcoming publication.

\section*{References}

\begin{enumerate}

\item U. Amaldi, W. de Boer, and H. Furstenau, Phys. Lett. {\bf B260},
447 (1991); P. Langacker and M.X. Luo, Phys. Rev. {\bf D 44}, 817
(1991); J. Ellis, S. Kelley, and D.V. Nanopoulos, Phys. Lett.
{\bf B260}, 131 (1991).
\item H. Fritzsch, Phys. Lett. {\bf 70B}, 436 (1977); Nucl. Phys. {\bf
B155}, 189 (1979).
\item CDF Collaboration, F. Abe et. al., Phys. Rev. Lett. {\bf 73}, 225
(1994).
\item K.S. Babu and Q. Shafi, Phys. Rev. {\bf D 47}, 5004 (1993); Y.
Achiman and T. Greiner, Phys. Lett. {\bf B329}, 33 (1994).
\item H. Georgi and C. Jarlskog, Phys. Lett. {\bf 86B}, 297 (1979).
\item J. Harvey, P. Ramond and D. Reiss, Phys. Lett. {\bf B92}, 309
(1980); Nucl. Phys. {\bf B199}, 223 (1982).
\item X.G. He and W.S. Hou, Phys. Rev. {\bf D41}, 1517 (1990)
\item S. Dimopoulos, L.J. Hall and S. Raby, Phys. Rev. Lett. {\bf 68},
752 (1992); Phys. Rev. {\bf D 45}, 4192 (1992).
\item H. Arason, D. Castano, E.J. Pirad and P. Ramond, Phys. Rev.
{\bf D 47}, 232 (1993).
\item V. Barger, M.S. Berger, T. Han and M. Zralek, Phys. Rev. Lett.
{\bf 68}, 3394 (1992).
\item P. Ramond, R.G. Roberts and G.G. Ross, Nucl. Phys. {\bf B406}, 19
(1993).

\item D. Lewellen, Nucl. Phys. {\bf B337}, 61 (1990); A. Font, L.
Ibanez, and F. Quevedo, Nucl. Phys. {\bf B345}, 389 (1990).
\item S. Chaudhuri, S. Chung and J. Lykken, FERMILAB-PUB-94/137-T
(1994).
\item G. Cleaver, Ohio State Preprint OHSTPY-HEP-T-94-007 (1994).
\item D. Lee and R.N. Mohapatra, Maryland Preprint UMD-PP-94-166 (1994);
\item S. Dimopoulos and F. Wilczek, Report No. NSF-ITP-82-07, August 1981
(unpublished); R.N. Cahn, I Hinchliffe and L. Hall, Phys. Lett.
{\bf 109B}, 426 (1982).
\item K.S. Babu and S.M. Barr, Phys. Rev. {\bf D 48}, 5354 (1993);
Phys. Rev. {\bf D 50}, 3529 (1994).
\item K.S. Babu and S.M. Barr, Bartol Preprint BA-94-45 (1994).
\item S.M. Barr, Phys. Rev. Lett. {\bf 64}, 353 (1990);
Z. Berezhiani and R. Rattazzi, Nucl. Phys. {\bf B 407}, 249 (1993);
G. Anderson,
S. Dimopoulos, L. Hall, S. Raby and G.D. Starkman, Phys. Rev. {\bf D 49},
3660 (1994).
\end{enumerate}

\newpage
\noindent {\bf Table. 1.}  The matter superfields along with their discrete
charge assignments.  The subscript in $Z$ and $\overline{Z}$ stand for
a $Z_2$ symmetry under which all other fields are even.
\begin{center}
\begin{tabular}{|c||c||c|}  \hline
Superfield & $Z_8\times Z_{16} \times Z_{16}\times Z_{4}$ \\
\hline
$\psi_H({\bf 16})$ &  (1,0,0,0) \\
$\bar{\psi}_H(\overline{\bf 16})$ & (7,2,0,0)\\
$H_1({\bf 10})$ & (6,0,0,0) \\
$H_2({\bf 10})$ & (2,12,0,0)\\
$A({\bf 45})$ & (0,4,0,0) \\
$A^{\prime}({\bf 45})$ & (0,12,0,2) \\
$Y({\bf 1})$ & (0,1,0,0)\\
$X({\bf 1})$ & (0,14,0,0)\\
$Z({\bf 1})$ & (0,4,0,0)$_-$ \\
$\bar{Z}({\bf 1})$ & (0,12,0,0)$_-$\\
$\psi_1({\bf 16})$ & (7,4,6,1)\\
$\psi_2({\bf 16})$ & (1,0,0,1) \\
$ \psi_3({\bf 16})$ & (0,0,1,1)\\
$S_1({\bf 1})$ & (2,0,14,2)\\
$S_2({\bf 1})$ & (14,4,14,2) \\
$S_3({\bf 1})$ & (13,4,15,2)\\ \hline
\end{tabular}
\end{center}

\end{document}